\documentclass[namedreferences]{kluwer}
\usepackage[]{graphicx}
\begin{document}
\begin{article}
\begin{opening}
\title{Probing the obscuring medium around active nuclei using masers: The case of 3C\,403}    

\author{A. \surname{Tarchi$^{1,2}$}}
\author{C. \surname{Henkel$^{3}$}}
\author{M. \surname{Chiaberge$^{1}$}}
\author{K. M. \surname{Menten$^{3}$}}
\author{A. \surname{Brunthaler$^{4}$}}
\author{L. \surname{Moscadelli$^{2}$}}
\institute{$^{1}$Istituto di Radioastronomia, CNR, Via Gobetti 101, 40129 Bologna, Italy}
\institute{$^{2}$INAF-Osservatorio Astronomico di Cagliari, Loc. Poggio dei Pini, Strada 54, 09012 Capoterra (CA), Italy}
\institute{$^{3}$Max-Planck-Institut f{\"u}r Radioastronomie, Auf dem H{\"u}gel 69, D-53121 Bonn, Germany}                               
\institute{$^{4}$JIVE, Joint Institute for VLBI in Europe, PO Box 2, 7990 AA Dwingeloo, The Netherlands}



\runningtitle{Obscuring medium around active nuclei: The case of 3C\,403}
\runningauthor{Tarchi et al.}

\begin{ao}
INAF-Osservatorio Astronomico\\
Loc.\ Poggio dei Pini, Strada 54\\
09012 Capoterra (CA)\\
Italy
\end{ao} 


\begin{abstract} 
We report the first detection of a water megamaser in a radio-loud galaxy, 3C\,403, and present a follow-up study using the VLA. 3C\,403 has been observed as a part of a small sample of FR\,II galaxies with evidence of nuclear obscuration. The isotropic luminosity of the maser is $\sim$ 1200 {\hbox{L$_{\odot}$}}. With a recessional velocity of c{\it{z}} $\sim$ 17680 km\,s$^{-1}$ it is the most distant water maser so far reported. The line arises from the densest ($>$ 10$^{8}$ cm$^{-3}$) interstellar gas component ever observed in a radio-loud galaxy. Two spectral features are identified, likely bracketing the systemic velocity of the galaxy. Our interferometric data clearly indicate that these arise from a location within 0.1$^{\prime\prime}$ ($\approx$ 110 pc) from the active galactic nucleus. We conclude that the maser spots are most likely associated with the tangentially seen parts of a nuclear accretion disk, while an association with dense warm gas interacting with the radio jets cannot yet be ruled out entirely.
\end{abstract}

\keywords{3C\,403, active galaxies, masers}



\end{opening}
\section{Introduction}
So far, water megamasers have been detected in radio-quiet active galactic nuclei (AGN; for a definition of radio-quiet AGN, see e.g.\ \opencite{kellerman89}), mostly in Seyfert\,2 and LINER galaxies. Under the assumption that all 
megamasers are associated with molecular material orbiting around the central engine (e.\ g.\ \opencite{miyoshi95}) or interacting with the nuclear jet(s) of the host galaxy (e.\ g.\ \opencite{claussen98}) and that their amplification is unsaturated (i.e. the maser intensity 
grows linearly with the background radio continuum), one should expect a much higher detection rate 
in radio-loud AGN than is actually observed. This fact should be especially true for the radio-loud AGN classified as narrow-lined FR\,IIs for which the AGN unification scheme (e.g.\ \opencite{urry95}) requires the presence of geometrically and optically thick obscuring structure. 

Samples of radio galaxies belonging to the broad-lined and narrow-lined FR\,II
class have been recently observed with the Effelsberg telescope (Tarchi et al., in prep; Lara et al., priv.\ comm.). 
As in the case of FR\,Is \cite{henkel98}, no maser detections have been obtained.

\section{Sample selection}

\begin{figure}
\centerline{\includegraphics[width=20pc]{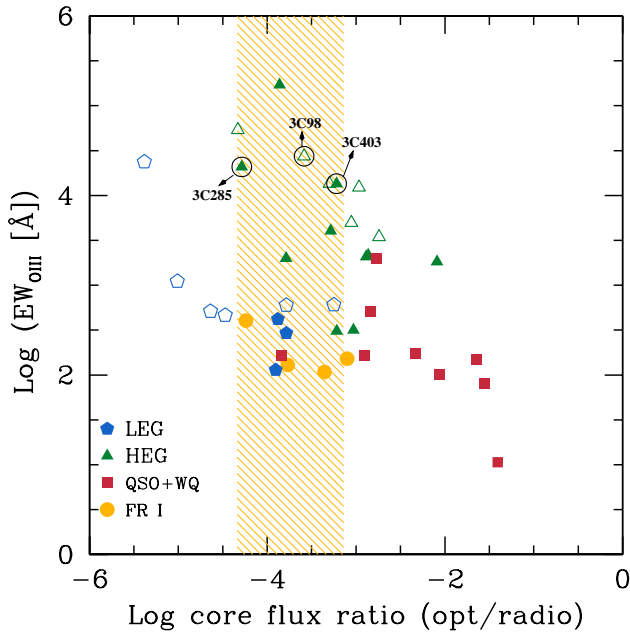}}
\caption{Equivalent width of the [O{\sc iii}] emission line, measured 
with respect to the central compact core emission, is plotted vs. the ratio between the optical central compact core to radio core flux (for details, see also Chiaberge et al., 2002). LEG: Low excitation galaxy; HEG: High excitation galaxy. The targets of our sample with their respective names are those within circles.}
\label{chiab}
\end{figure}

Our sample comprises all nearby ($z<0.1$) 3 FR\,IIs spectrally classified as High Excitation Galaxies  
(HEGs, \opencite{jackson97}) with nuclear equivalent widths of the [O{\sc iii}]$\lambda$5007 emission 
line EW([O{\sc iii}])$>10^{4}$ {\AA} (Fig.\,\ref{chiab}). A high 
value for the nuclear EW([O{\sc iii}]) in HEGs has been interpreted as a hint for the obscuration of the central ionizing 
continuum source \cite{chiaberge02}. In these sources the nuclear ionizing continuum would be 
obscured to our line-of-sight and only a small fraction of the emission is seen through scattered light.
Therefore, our selection criteria provide us with a sample of galaxies with both high radio flux densities and 
nuclear obscuration of the central ionizing source.

\section{Observations and data reduction} 

Observations of the $6_{16} - 5_{23}$ transition of H$_2$O (rest frequency: 22.235 GHz) were carried out with 
the 100-m telescope of the MPIfR at Effelsberg\footnote{The 100-m telescope at Effelsberg is operated by the 
Max-Planck-Institut f{\"u}r Radioastronomie (MPIfR) on behalf of the Max-Planck-Gesellschaft (MPG).} in January 
and March 2003. The beam width (HPBW) was 40$^{\prime\prime}$. Flux calibration was obtained by measuring W3(OH) (see \opencite{mauer88}). 
Gain variations of the telescope as a function of elevation were taken into account (Eq.\,1 of \opencite{gallimore01}). The pointing accuracy was better than 10$^{\prime\prime}$. 

The follow-up VLA\footnote{The National Radio Astronomy Observatory is a facility of the National Science Foundation operated under cooperative agreement by Associated Universities, Inc.} A-array observations of the detected maser in 3C\,403 were performed in July 2003 with two IFs and a bandwith of 12.5 MHz each centered on one of the two maser features. Using 32 spectral channel a velocity resolution of $\sim$ 6\,km\,s$^{-1}$ was reached. The beam width (HPBM) was $\sim$ 0.1$^{\prime\prime}$ and the total on-source observation time was about $\sim$ 8 hours.
 
All data were reduced using standard procedures belonging either to the GILDAS or the AIPS software packages.

\section{Results} 

\begin{figure}
\centerline{\includegraphics[width=20pc]{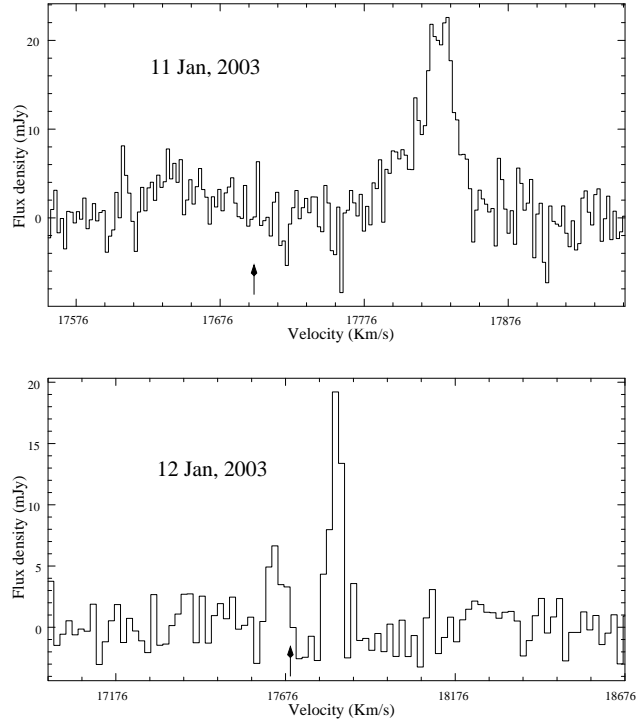}}
\caption{Maser lines in 3C\,403. {\it Upper panel} Spectral resolution of 78 kHz = 1.15\,km\,s$^{-1}$. {\it Lower panel} Spectral resolution of 1.25 MHz = 17.8\,km\,s$^{-1}$. The arrow marks the nominal systemic velocity of the galaxy = 17688\,km\,s$^{-1}$.}
\label{2spec}
\end{figure}

The maser spectra of 3C\,403, taken in January with the Effelsberg telescope, are shown in Fig.\,\ref{2spec}. The profile is composed of two main
components (asymmetrically) bracketing the nominal systemic velocity of the galaxy (c$z$ = 17688\,km\,s$^{-1}$, see next section): the 
stronger one has a velocity of c$z$ = 17827$\pm$1\,km\,s$^{-1}$, a width of 31$\pm$2\,km\,s$^{-1}$, and a flux density peak of 23$\pm$3\,mJy; 
the weaker one has a velocity of c$z$ = 17644$\pm$5\,km\,s$^{-1}$, a width of 53$\pm$8\,km\,s$^{-1}$, and a peak flux density of 4.0$\pm$0.5\,mJy. Using a distance of 235\,Mpc ($H_{\rm 0}$ = 75\,km\,s$^{-1}$\,Mpc$^{-1}$), the total isotropic 
luminosities are 950$\pm$140 and 280$\pm$55\,\hbox{L$_{\odot}$} for the main components, respectively. Hence, the maser in 3C\,403 is the most distant and one of the most luminous water maser so far reported (the most luminous maser, with 6000\,\hbox{L$_{\odot}$}, and, before this discovery also the most distant one, is in TXFS2226-184; \opencite{koek95}) 

\begin{figure}
\centerline{\includegraphics[width=30pc]{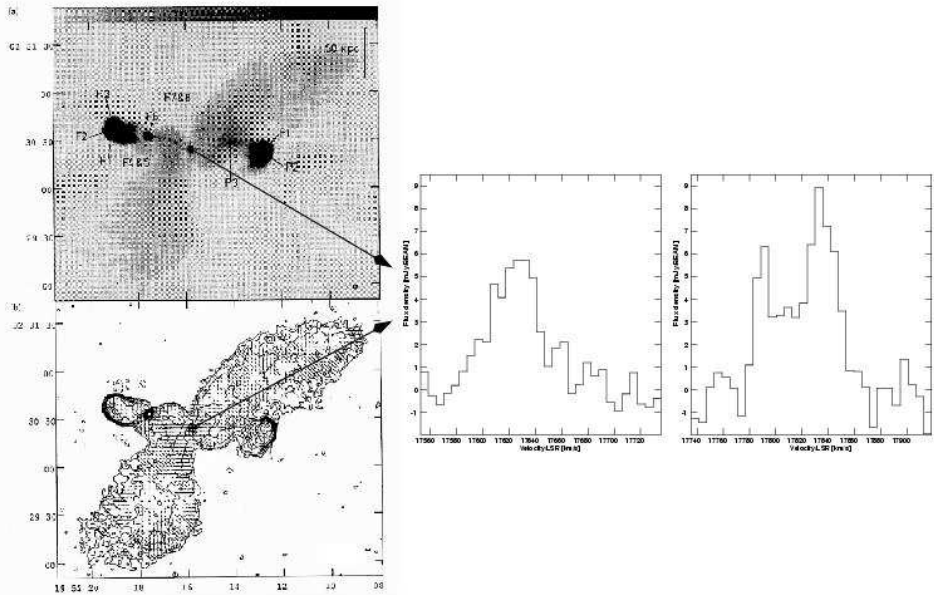}}
\caption{{\it Left-hand side} VLA 8.4 GHz maps in grey-scale (top) and contours (bottom) of 3C\,403 at a resolution of $\sim$ 2.5$^{\prime\prime}$ (FWHM). For details, see Black et al. (1992). {\it Right-hand side} The two main water maser features in 3C\,403 observed with the VLA A-array. Both lines arise from a location within 0.1$^{\prime\prime}$ ($\approx$ 110 pc) from the active galactic nucleus (RA${\rm _{B1950}}$ = 19$^{\rm h}$ 52$^{\rm m}$ 15.8$^{\rm s}$; Dec${\rm _{B1950}}$ = 02$^{\rm \circ}$ 30$^{\rm \prime}$ 24$^{\rm \prime\prime}$).}
\label{vla}
\end{figure}

The outcome of the VLA observations (Fig.\,\ref{vla}; right-hand side) confirms the presence of the two main maser features, and indicates that the maser emission is unresolved and arise from a location within 0.1$^{\prime\prime}$ ($\approx$ 110 pc) from the active galactic nucleus. The strength of the weaker line is consistent with the single-dish result, while the stronger line is much weaker than that obtained with Effelsberg. Because of the unresolved nature of the emission it is unlikely that the loss of flux density in the feature is due to extended emission resolved-out because of the higher resolution. More probably, the cause is a flare-down of the component due to strong variability, a phenomenon already well known to exist in water masers. 

\section{Discussion and conclusions} 

The standard unified scheme of AGN requires an obscuring region, possibly containing molecular gas that surrounds 
the central engine and that effectively shields the inner few parsecs from view, if the radio axis lies close to 
the plane of the sky \cite{antonucci93}. In the innermost part, at radii up to some tenths of a parsec, 
this material is likely to form a rapidly rotating accretion disk around a central supermassive black hole. At 
larger distances (up to about 50-100 pc) the atomic and molecular gas is possibly distributed in a toroidal structure 
providing obscuration of the central regions to particular lines-of-sight.

Indeed, very few direct 
detections of molecular gas in radio-loud galaxies have been reported so far ($\rm H_{2}$ in Cygnus\,A: \opencite{evans99}; CO in 3C\,293: \opencite{wilman00}). The detection discussed here strongly favors the presence 
of molecular material near the central engines of at least some FR\,IIs. Past negative results in molecular line surveys 
may also be a consequence of observational sensitivity limits (such a possibility was also mentioned by \opencite{henkel98}). Furthermore, because of the high gas densities required for H$_{2}$O masers to operate ($>$10$^{8}$\,cm$^{-3}$; 
e.\ g.\ \opencite{elitzur89}), our detection represents the densest interstellar gas component ever observed 
in a radio-loud galaxy (typical values for molecular gas densities range between $\sim 10^{3}$ and $ \sim 10^{5}$ cm$^{-3}$; 
e.\ g.\ \opencite{wilman00}). 

\subsection{Accretion disk or jet interaction?}

Because of the large uncertainty of the systemic velocity\footnote{Velocities derived from optical emission lines may be uncertain or biased by motions of the emitting gas (e.g. \opencite{morganti01}). From the rotation curve measured by \inlinecite{baum90}, we deduce that these uncertainties in 3C\,403 are $<$ 100\,km\,s$^{-1}$.} ($V_{\rm sys}$) of 3C\,403, the following discussion and (preliminary) conclusions are based on the assumption that $V_{\rm sys}$ is placed (as indicated in Fig.\,\ref{2spec}) between the two main H$_{2}$O maser components.

As mentioned in Sect.\,1, interferometric studies of H$_2$O megamasers have shown that the emission is either associated 
with a nuclear accretion disk (for NGC\,4258, see e.g.\ \opencite{miyoshi95}) or with the radio-jets interacting 
with dense molecular material near the center (for Mrk\,348, see \opencite{peck03}). 

The interferometric observation performed with the VLA in its A configuration confirms that also the megamaser in 3C\,403 has a nuclear origin.

Only VLBI interferometric observations at milliarcsecond resolution (spatial scales of $\approx$ 1 parsec) 
will allow us to determine (or to provide an upper limit to) 
the extent of the emission and to pinpoint the exact location of the H$_2$O emitting region(s) in order to see if the maser is associated with an accretion disk or radio jets. Nevertheless, a 
qualitative discussion is possible on the basis of the single-dish spectra of Fig.~\ref{2spec} and the new VLA observation.

{\it {Accretion disk}}: this scenario is particularly supported by the expected almost edge-on orientation of the nuclear obscuring layer (see \opencite{chiaberge02}). As in the case of the Seyfert galaxy 
NGC~4258 (e.g.\ \opencite{miyoshi95}), the spectrum should then show three distinct groups of features: one 
centered at the systemic velocity of the galaxy (originating along the line of sight to the nucleus) and two groups 
symmetrically offset from the systemic velocity, arising from those parts of the disk that are viewed tangentially. 
If the latter are the two lines we are observing, the rotational velocity of the disk is $\sim$100\,km\,s$^{-1}$ (which is, we have to point out, very small when compared with that derived for NGC\,4258). In 3C\,403 the systemic lines seem instead to be missing. To explain this fact, partly following the hypothesis proposed by \inlinecite{henkel98}, we could argue that the circumnuclear disk is actually a {\it {thin rotating ring}} of the type modeled by \citeauthor{ponomarev94} (1994; their Fig.\,2), where masing gives a double-peaked profile only.

{\it {Jet interaction}}: The handful of bright knots visible in the radio images shown in Fig.\,\ref{vla} (left-hand side) hints at the presence of jets 
interacting in several regions with the interstellar medium. The profile of the spectrum does not contradict a `jet-origin' 
of the detected maser emission in 3C\,403. If we assume a symmetric molecular distribution, the two observed features 
could be interpreted as the red-shifted and blue-shifted counterparts of the maser line, arising from opposite jets close to the core.

\end{article}
\end{document}